\documentclass[prx,superscriptaddress,twocolumn]{revtex4-1}
\usepackage{times,amsmath}
\usepackage{epsfig}
\usepackage{color}
\usepackage{longtable}
\usepackage{graphicx}
\usepackage{dcolumn}
\usepackage{bm}
\usepackage{tabularx}
\usepackage{hyperref}
\usepackage{multirow}
\hypersetup{citecolor=black, filecolor= black, linkcolor= black, urlcolor= black}

\begin{document}
\title{First-principles investigation of Sc-III/IV under High Pressure}

\author{Sheng-Cai Zhu}
\affiliation{Department of Physics and Astronomy, High Pressure Science and Engineering Center, University of Nevada, Las Vegas, NV 89154, USA}

\author{Xiao-Zhi Yan}
\affiliation{Academy for Advanced Interdisciplinary Studies, and Department of Physics, Southern University of Science and Technology, Shenzhen, 518055, China}

\author{Scott Fredericks}
\affiliation{Department of Physics and Astronomy, High Pressure Science and Engineering Center, University of Nevada, Las Vegas, NV 89154, USA}

\author{Yan-Ling Li}
\email{ylli@jsnu.edu.cn}
\affiliation{School of Physics and Electronic Engineering, Laboratory for Quantum Design of Functional Materials, Jiangsu Normal University, Xuzhou, 221116, China}

\author{Qiang Zhu}
\email{qiang.zhu@unlv.edu}
\affiliation{Department of Physics and Astronomy, High Pressure Science and Engineering Center, University of Nevada, Las Vegas, NV 89154, USA}

\date{\today}
\begin{abstract}

Using \textit{ab initio} evolutionary structure prediction method in conjunction with density functional theory, we performed a systematic investigation on the structural transition of elemental scandium under pressure up to 250 GPa. Our prediction successfully reproduced several allotropes which have been reported in the previous literature, including the Sc-I, Sc-II and Sc-V. Moreover, we observed a series of energetically degenerate and geometrically similar structures at 110-195 GPa, which are likely to explain the unsolved phases III and IV reported by Akahama [Phys. Rev. Lett., \textbf{94}, 19, 195503, (2005)]. A detailed comparison on powder X-ray diffraction pattern (PXRD) suggested that the $Ccca$-20 phase may account for the observed Sc-III, while Sc-IV is likely to be explained by a mixture of multiple energetically competing structures. We also used the candidate Sc-III structure as the model system to explore its superconducting behavior under pressures between 80-130 GPa. The predicted superconducting transition temperature $T_\textrm{c}$ values are in satisfactory agreement with previous experimental results.

\end{abstract}
\maketitle

\section{Introduction}
Elemental solids are the most fundamental cases for scientific studies on materials \cite{mcmahon-2006-review, bridgman1964collected}. External pressure can effectively squeeze the crystal packing, alter the electronic configuration and thus trigger the phase transition. Knowing the atomic structures is the key to understand their properties under high pressure \cite{mcmahon-2006-review}. Searching for new allotropes under high pressure has been a long term interest for scientists \cite{young1991phase,degtyareva2004-Ga-structural,arapan2008prediction-Ca-}. To date, many new structures and intrigue properties have been discovered for the advance in high-pressure techniques. For example, some simple metals, such as Li \cite{hanfland2000new-Li,lv2011predicted-Li,matsuoka2009direct-Li,struzhkin2002superconductivity-Li,marques2011crystal-Li} and Na \cite{ma2009transparent-Na,gregoryanz2008-Na,mcmahon2007structure-Na}, transform to semi-metallic, semiconducting and even insulating phases under high pressure \cite{rousseau2011exotic}. 

Sc, as the first 3$d$-transition-metal, has attracted a special interest \cite{debessai2008comparison-Tc,ormeci2006first-electricstructure,bose2008linear-Sc-elephonon,kong2011first-calculation}. In the past, scandium was often grouped with the rare-earth metals in the IIIB group since its mechanical, physical, and chemical properties are similar to those of Y, La, Pr, Nd, etc \cite{grosshans1982high}. Previous studies showed that group IIIB metals exhibit successive pressure-induced phase transitions \cite{debessai2008comparison-Tc,samudrala2012structural-Y,chesnut1999structural-Y,olijnyk2004lattice-Pu,samudrala2011high-Er}, due to the electron transfer known as $s \rightarrow d$ transition under pressure \cite{zeng1997structure-s-d-,skriver1985crystal-s-d--,morozova2015features-Sc-Y-Eu,debessai2009pressure-s-d}. These phase transitions follow a systematic sequence of hexagonal close packing (hcp) $\rightarrow$ Sm-type structure $\rightarrow$ double hexagonal close packing (dhcp)$\rightarrow$ face centered cubic (fcc) $\rightarrow$ double face centered cubic (dfcc) \cite{johansson1975generalized-Y-Lu-,nixon2007-calculations-Tc,debessai2008comparison-Tc}. Sc was suggested to follow the same series of phase transitions under pressure, as found in Y and La \cite{vohra1982high,olijnyk2006unusual-Sc-work}. However, two recent high quality powder X-ray diffraction (PXRD) studies showed that the first high pressure structure of Sc, known as Sc-II, stable between 23 and 104 GPa, adopt an incommensurate structure (IC) consisting of two interpenetrating sublattices along the crystallographic $c$ axis \cite{mcmahon2006different-Sc-II-powder,fujihisa2005incommensurate-highquality-XRD}, making Sc distinct from other group IIIB metals \cite{akahama2005new-240GPa,mcmahon2006different-Sc-II-powder}. This also provides a first example of IC structure observed in non main-group elements \cite{mcmahon2004incommensurate-maingroup}. The intriguing structural complexity has stimulated a series of experimental and theoretical works on scandium \cite{molodets2007electrophysical-shock,briggs2017ultrafast-shock,olijnyk2006unusual-Sc-work,ormeci2006first-electricstructure,PRL-Sc-II-2009,olijnyk2006unusual-Sc-work}.

\begin{figure}
\epsfig{file=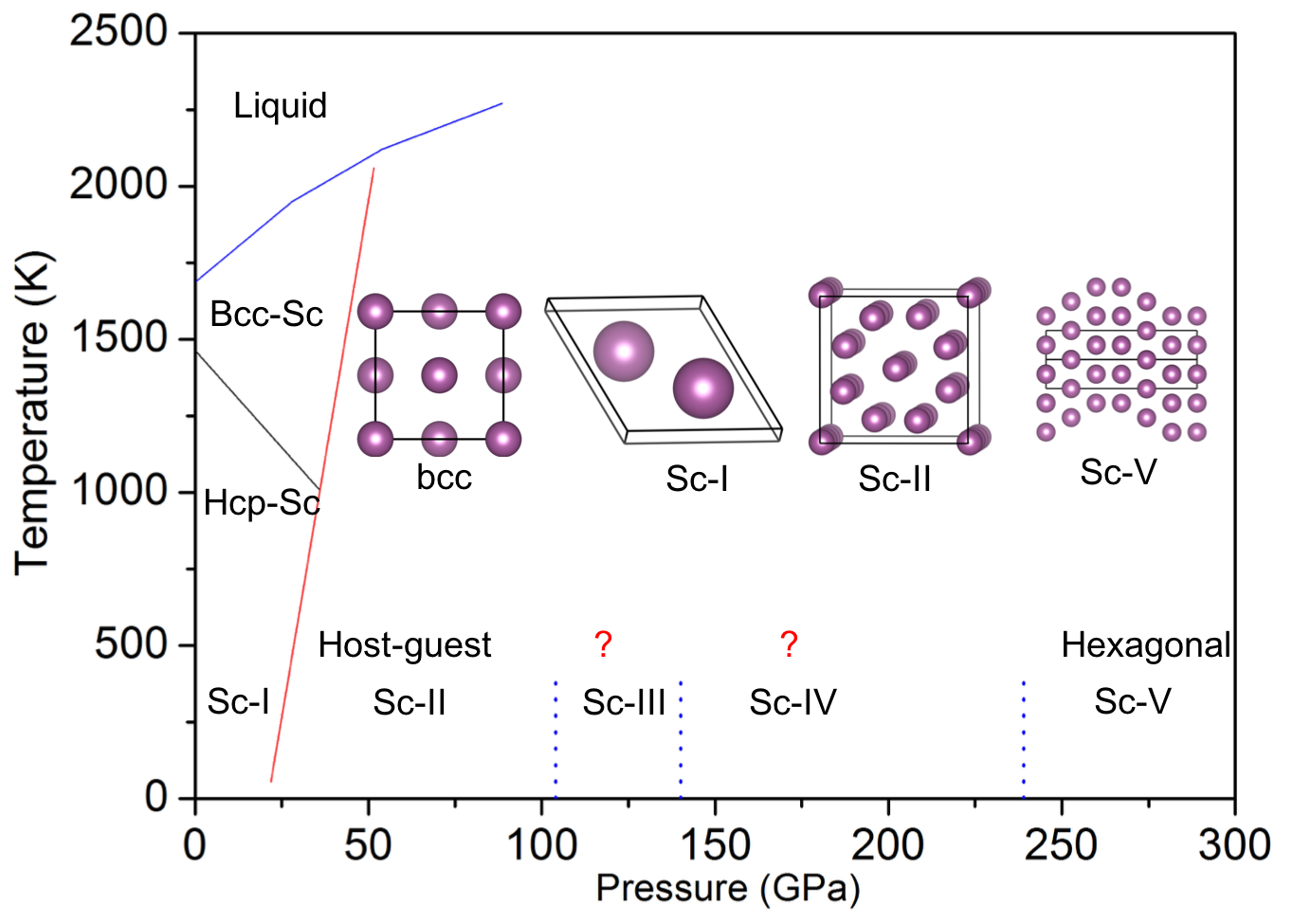, width=0.45\textwidth}
\caption{\label{PhaseDiagram} {The pressure-temperature phase diagram of scandium reproduced from references \cite{akahama2005new-240GPa,briggs2017ultrafast-shock}. The inset shows the atomic structures of bcc, hcp (Sc-I), Host-guest (Sc-II) and hexagonal (Sc-V) except Sc-III and Sc-IV.}}
\end{figure}

Experimentally, Sc was found to exhibit resistant anomalies at about 17 GPa \cite{wittig1979superconductivity-17GPa}, and it becomes superconductor at 20 GPa. The superconducting transition temperature ($T_\textrm{c}$) rapidly soars up when increasing the external pressure \cite{hamlin2007pressure-experiment-Tc,vohra1982high}. Recently, it was found that scandium reaches the highest $T_\textrm{c}$ of 19.7 K at 107 GPa and then drops to about 8 K under further compression \cite{debessai2008comparison-Tc}. The sudden drop of $T_\textrm{c}$ at 107 GPa is believed to be triggered by the structural phase transition. Using the monochromatic synchrotron PXRD technique, Sc was found to undergo four stages of structural transitions, i.e. Sc-I ($P6_3/mmc$) $\rightarrow$ Sc-II ($I4/mcm$) $\rightarrow$ Sc-III (unsolved) $\rightarrow$ Sc-IV (unsolved) $\rightarrow$ Sc-V ($P6_122$), at around 23, 104, 140 and 240 GPa \cite{zhao1996evidence-Sc-23GPa,akahama2005new-240GPa}, respectively (see Fig. \ref{PhaseDiagram}). Unfortunately, two structures (Sc-III and Sc-IV) are still unclear \cite{riva2016scandium}. As Akahama reported \cite{akahama2005new-240GPa}, these structures may contain a large number of atoms in the unit cell, according to the observed complex PXRD profiles. Due to the lack of the atomic models of Sc-III and IV, the electron-phonon coupling characteristic of Sc at high pressure beyond 107 GPa is still a mystery.

\begin{figure*}
\epsfig{file=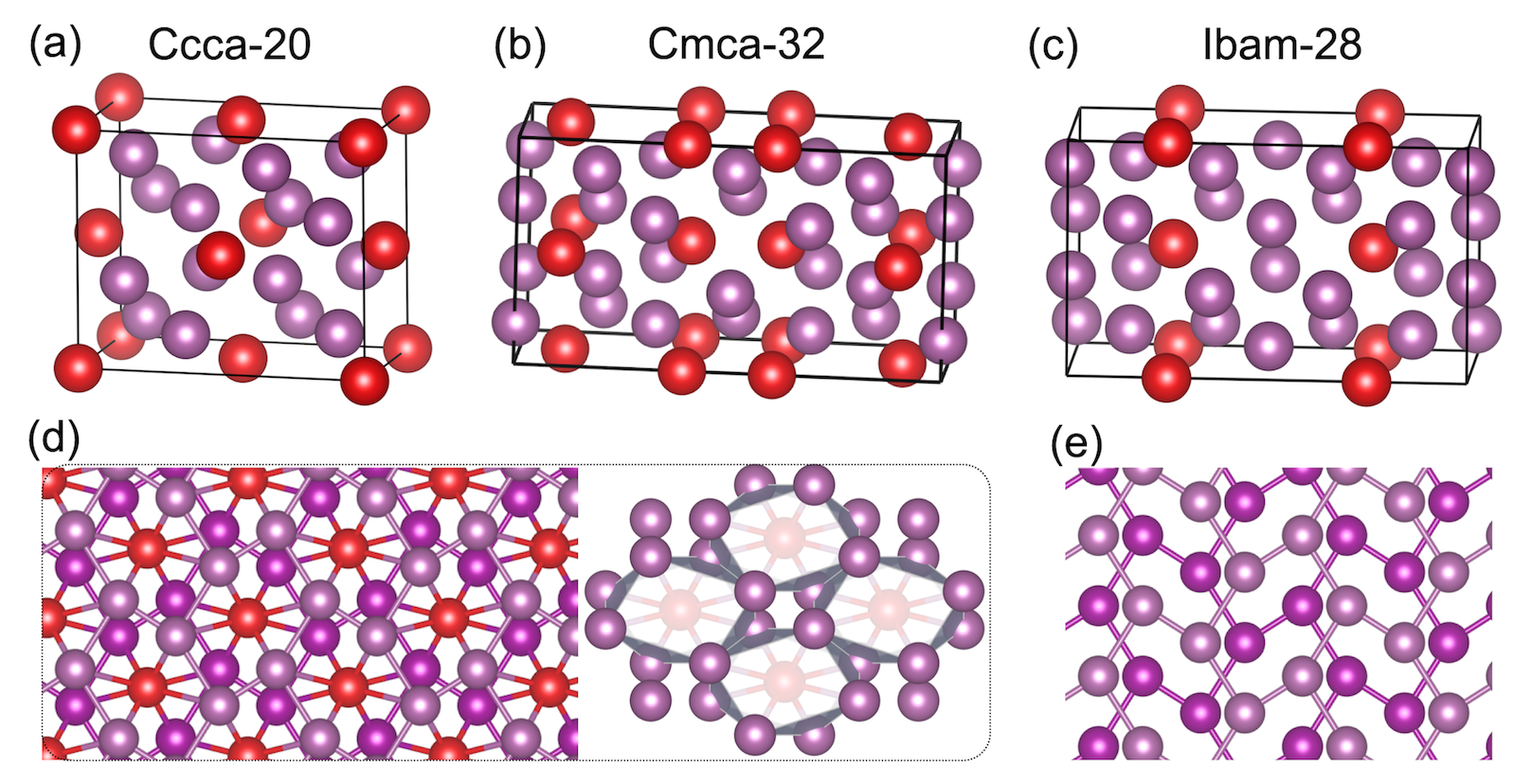, width=0.8\textwidth}
\caption{\label{structure} {The atomic structures of $Ccca$-20 (a), $Cmca$-32 (b), $Ibam$-28 (c), (d) illustrates the model with distorted-hexagon layer with intercalated atoms (left) and its polyhedron representation (right), (e) illustrates the hexagon layers without intercalated atoms. The atoms belonging to the distorted hexagon framework are denoted by purple spheres, while the intercalated atoms are denoted by red spheres.}}
\end{figure*}

Here, we explored the high-pressure effects on Sc by using the \textit{ab initio} evolutionary structure prediction method USPEX \cite{Oganov-JCP-2006, Lyakhov-CPC-2013}. Through an extensive crystal structure search, we found a series of structures which are both energetically and geometrically degenerate at 110-195 GPa. Interestingly, all these structures possess two types of atoms: 1) one builds the layered distorted-hexagon framework, 2) the other can be explained as the intercalated atoms between the distorted-hexagon layers. By comparing the simulated PXRD profiles with previous experiment data, we suggest that $Ccca$-20 (no. 68) structure (ground state at pressures between 75 and 160 GPa) as the candidate model for Sc-III. While, Sc-IV may be explained by a mixture of two metastable allotropes, including $Ibam$-28 (no. 72) and $Ibam$-12 (no. 72), rather than the ground state $Cmca$-32 (no. 64) structure at pressures between 160 and 195 GPa. We also used the candidate Sc-III structure as the model system to explore its superconducting behavior. The predicted pressure dependence of $T_\textrm{c}$ is in satisfactory agreement with previous experimental results \cite{debessai2008comparison-Tc}.

\section{Computational Methods}
Based on evolutionary structure prediction method USPEX code \cite{Oganov-JCP-2006, Lyakhov-CPC-2013} in conjunction with first-principles calculations, we performed several runs at 0, 30, 110, 150, 180, and 250 GPa with no more than 32 atoms in the unit cell. During the structure search, the first generation of structures were created randomly, the worst structures (40\%) were discarded and the best structures from each generation were kept. Next generations were created by heredity, mutation and random generator operations. All structures optimization evolved over maximum of 40 generations. Each structure was optimized at the level of density functional theory (DFT) as implemented in the VASP code \cite{Vasp-PRB-1996}. The exchange-correlation functional was described by the generalized gradient approximation in the Perdew-Burke-Ernzerhof parameterization (GGA-PBE) \cite{PBE-PRL-1996}, and the energy cutoff of the plane wave was set as 1000 eV. The geometry convergence criterion was set as 0.001 eV/{\AA} for the maximal component of force and 0.01 GPa for stress. The Brillouin zone was sampled by uniform $\Gamma$-centered meshes with the reciprocal space resolution of 2$\pi$ $\times$ 0.03 {\AA$^{-1}$}. In order to check the dynamical stability of the candidate structures, we also carried out phonon calculations with the finite displacement method as implemented in Phonopy code \cite{Togo-PRB-2008}.

To explore the superconducting properties for the selected structures, we also calculated their superconducting properties by using the Quantum Espresso package \cite{QE} based on the projected augmented wave (PAW) potentials with cutoff energies of 100 Ry and 800 Ry for the wave functions and the charge density, respectively. The electronic band structure and density of states were computed with a $24 \times 24 \times$ 24 Monkhorst-Pack (MP) mesh. The electronic Brillouin zone (BZ) integration in the phonon calculation was based on a $16 \times 16 \times 16$ of Monkhorst-Pack k-point meshes. The dynamic matrix was computed based on a $4 \times 4 \times 4$ mesh of phonon wave vectors. The electron-phonon coupling was convergent with a finer grid of $24 \times 24 \times 24$ k points and a Gaussian smearing of 0.01 Ry.

\section{Results and Discussions}

\subsection{The Phase Diagram of Sc as a function of Pressure}
First, we found the hcp structure ($P6_3/mmc$, no. 194) is the most stable structure at 0 GPa, and the $I4/mcm$ (no. 108) structure as the ground state at 30 GPa. The $I4/mcm$ structure is believed to be the simplest approximates of the IC model of Sc-II. The excellent agreements between theory and experiment encouraged us to explore the high pressure effects further. At 110 GPa, our simulation found the orthorhombic $Ccca$ (no. 68) structure with 20 atoms in the conventional cell (as shown in Fig. \ref{structure}a) has the lowest-enthalpy. Its lattice parameters at 120 GPa are $a$ = 7.8518 {\AA}, $b$ = 6.4520 {\AA}, $c$ = 4.4536 {\AA}. In this structure, there are two sets of atomic sites, one in the general Wyckoff position 16i sites at (0.1435, 0.6427, 0.1288), and the other in the special Wyckoff position 4a sites at the origin (0, 0, 0). The atoms in the 16i sites build the close packed layer based on distorted hexagons, in which the 2/3 of Sc-Sc intralayer distances are 2.55 {\AA} and the remaining 1/3 of distances are 2.12 {\AA}. The 4a sites are occupied by the intercalated atoms between the adjacent distorted-hexagon layers. In the conventional unit cell, each hexagon layer contains 4 atoms (denoted as \textbf{A} layer) in a close packing manner, while each intercalated layer contains 2 atoms (denoted as \textbf{B} layer) in a loose pack manner. They are arranged periodically along the crystallographic $a$-axis, and we call this stacking sequence as 2A+1B+2A+1B. 

At pressures between 150 and 180 GPa, there exist two energetically competitive orthorhombic structures, i.e., $Cmca$ (no. 64) structure with 32 atoms per unit cell and $Ibam$ (no. 72) structure with 28 atoms per unit cell (as shown in Fig. \ref{structure}b-c). $Cmca$-32 is the most stable structure when pressure higher than 160 GPa. The lattice parameters at 150 GPa are $a$ = 12.2825 {\AA}, $b$ = 6.1338 {\AA}, $c$ =  4.4015 {\AA}. This structure has three different Wyckoff sites, 16g (0.8184, 0.1434, 0.8747), 8f (0.0000, 0.6609, 0.3900) and 8d (0.8961, 0.0000, 0.5000). Similar to $Ccca$-20, the Sc atoms at 16g and 8f sites build the distorted-hexagon layered framework, while Sc at 8d sites form the intercalated layers. $Ibam$-28 is marginally stable at 100-180 GPa, with the lattice parameters of $a$ = 4.3916 \AA, $b$ = 6.3096 \AA, $c$ = 10.5340 {\AA} at 150 GPa, and atoms occupying 16k (0.3691, 0.1517, 0.8529), 8j (0.1702, 0.8809, 0.0000) and 4a sites (0.5000, 0.5000, 0.2500). Comparing these two structures, they both contain six close packed distorted-hexagon layers. The difference lies in that intercalated atoms run every three hexagon layers in $Ibam$-28 (denoted as 3A+1B+3A+1B), while in $Cmca$-32 the intercalated atoms appear in every two and one hexagon layers (denoted as 2A+1B+1A+1B).  More interestingly, several energetically degenerate and geometrically similar structures can be constructed by changing the stacking sequence between \textbf{A} and \textbf{B} layers. As we will discuss in the following section, this phenomenon will lead to an infinite number of series of Sc allotropes. We also performed phonon calculations for all the three structures proposed in this work at different pressure conditions. The absence of imaginary frequencies in the phonon spectrum \cite{SI} confirms that they are all dynamically stable.

\begin{figure}
\epsfig{file=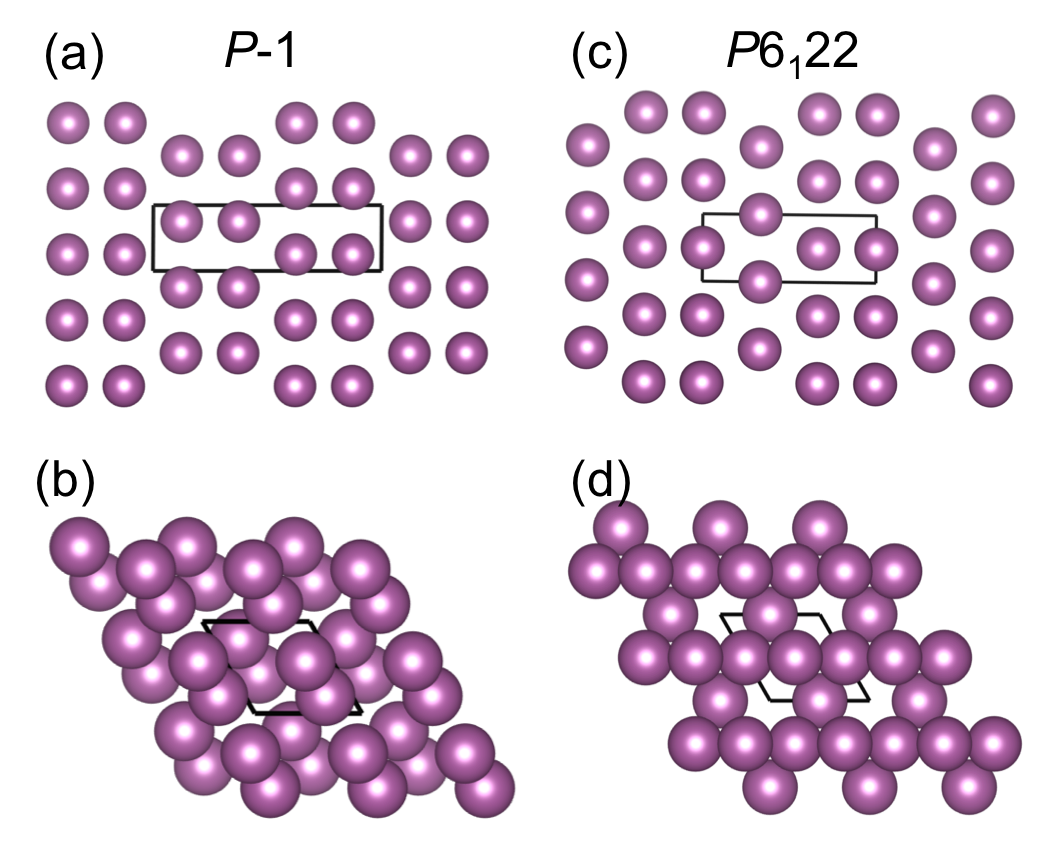, width=0.45\textwidth}
\caption{\label{structure2} {The atomic structure of $P$-1 side view (a) and top view (b), $P6_122$ side view (c) and top view (d).}}
\end{figure}

At 250 GPa, we found several structures based on the stacking of hexagon layers, while the intercalated layers disappear. The energetic of those structures are extremely close ($\sim$2 meV/atom), within the numerical error of DFT calculation. In the range of 200-290 GPa, the most stable structure is the $P$-1 structure (Fig. \ref{structure2}a-b), while the experimentally identified $P6_122$ \cite{akahama2005new-240GPa} is 3 meV/atom less stable than the $P$-1 structure at 250 GPa. Given that these two structures possess extremely different PXRD pattern (see Fig. S7 \cite{SI}), it is unlikely that the $P$-1 structure was present in experiment \cite{akahama2005new-240GPa}. However, this gap between experiment and theory may be explained by the missing of zero point energy (ZPE) and finite temperature effects in our calculation. Indeed, previous studies have found that ZPE can bring about 5 meV/atom for elemental Ca \cite{oganov2010-Ca}. We expect that the $P6_122$ structure would become favorable due to these effects. 

\begin{figure}
\epsfig{file=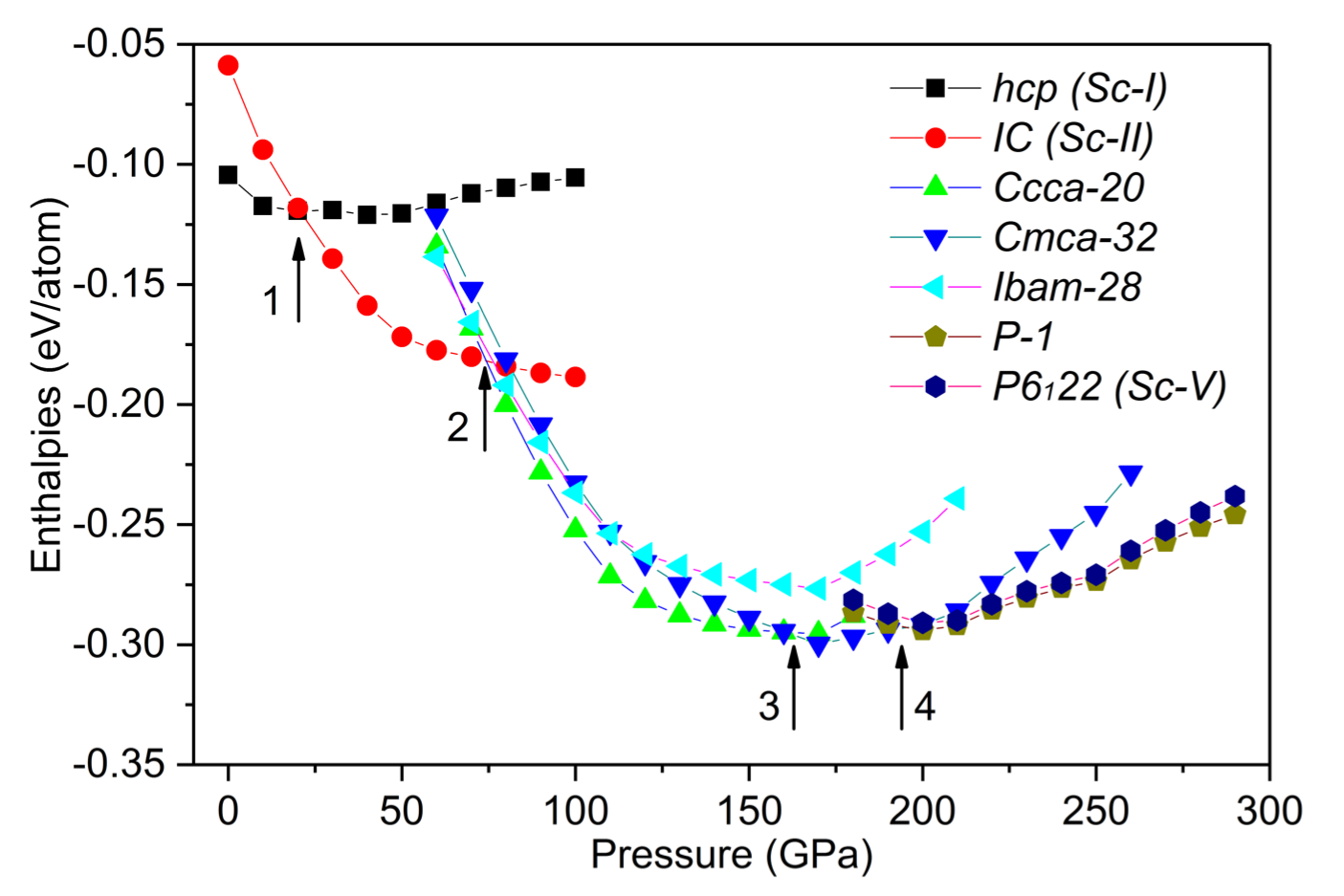, width=0.48\textwidth}
\caption{\label{enthalpies} {Enthalpies of the hcp, IC, $Ccca$-20, $Cmca$-32, $Ibam$-28, $P$-1 and $P6_122$ structures (relative to the bcc structure). The arrows indicate the four phase transition points. }}
\end{figure}

We further plotted the enthalpies for all relevant structures as a function of pressure in Fig. \ref{enthalpies} and Fig. S3 \cite{SI}. The ambient hcp phase remains stable up to 23 GPa, followed by $I4/mcm$ IC between 23 and 75 GPa, which is in excellent agreement with the experimental results \cite{akahama2005new-240GPa}. While $Ccca$-20 structure is calculated to be the most stable phase at pressure higher than 75 GPa. However, in experiment, both Akahama \textit{et al.} \cite{akahama2005new-240GPa} and Debessai \textit{et al.} \cite{debessai2008comparison-Tc} reported that the host-guest structure transits to Sc-III at about 104-107 GPa. This may be due to the fact that we only considered the $I4/mcm$ structure as the candidate model for Sc-II. It was report that under compression the incommensurate ratio $\gamma$ between host lattice ($c1$) and guest lattice ($c2$) undergoes a significant variation from 1.28 to 1.36 \cite{fujihisa2005incommensurate-highquality-XRD, PRL-Sc-II-2009}. Since this is not our focus, we do not include the modulation effects in our calculation. The $Cmca$-32 structure becomes most stable at 160 GPa, while $Ibam$-28 is energetically close in the entire pressure range studied in this work. At above 195 GPa, the hexagonal close layer packing $P$-1 structure has lowest-enthalpies. As described above, the ZPE correction ($\sim$ 5 meV/atom) would make $P6_122$ structure stable than $P$-1. We note that in experiment the Sc-IV to Sc-V transition takes place at about 240 GPa but our prediction is 195 GPa. The discrepancy may be due to kinetic reasons, or due to the limit of pseudopotential used in this study. Nevertheless, the prediction phase transition sequence in our study is overall in qualitative agreement with the experiment \cite{akahama2005new-240GPa}.

\subsection{Superconductivity of $Ccca$-20}

The superconducting behaviors for transition metals have been widely studied in the past. Unlike the simple $s$-metals, the $T_\textrm{c}$ of transition metals usually shows a highly nonlinear dependence as a function of pressure. Such complexity is attributed to the nature of $d$ electrons and also structural transitions under pressure \cite{debessai2008comparison-Tc}. As the first member in this group, the $T_\textrm{c}$ pressure dependence of Sc has been studied by several groups \cite{wittig1979superconductivity-17GPa,debessai2008comparison-Tc,bose2008linear-Sc-elephonon,kong2011first-calculation}. Debessai \textit{et al.} found that scandium reached the highest $T_\textrm{c}$ of 19.7 K at 107 GPa and then dropped to about 8 K under further compression till 123 GPa \cite{debessai2008comparison-Tc}. The sudden decrease of $T_\textrm{c}$ above 107 GPa is consistent with the phase transition pressure ($\sim$104 GPa) found by Mcmahon using the monochromatic synchrotron PXRD \cite{mcmahon2006different-Sc-II-powder}. In the past, an in-depth study on the superconducting behavior of Sc-III was prohibited due to the lack of structural model. Herein, we chose the most likely $Ccca$-20 as the model structure to explore its superconducting properties. We calculated its electronic band structure, density of states (DOS), phonon spectra and the Fermi surface at three different pressures, i.e. 80, 100 and 130 GPa. 

\begin{figure}
\epsfig{file=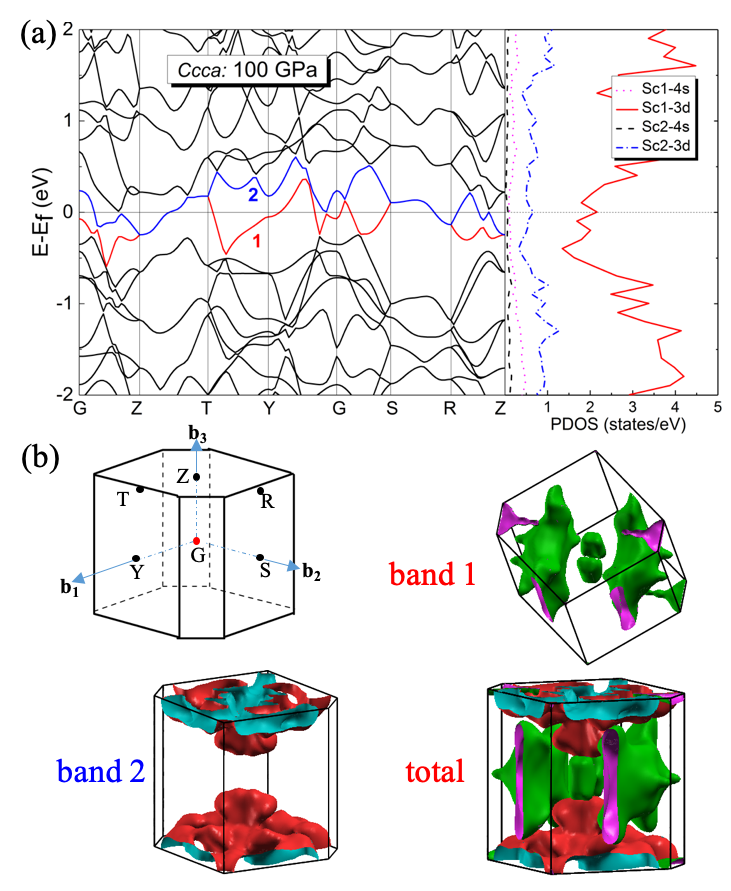, width=0.48\textwidth}
\caption{\label{electronicband} {(a) The electronic band structure along high symmetry lines of the Brillouin zone and projected DOS and the Fermi surface of Sc in the $Ccca$ phase calculated at 100 GPa. The energy bands crossing the Fermi level are labelled as 1 and 2, respectively.}}
\end{figure}

We found that the electron band structure of $Ccca$-20 does not notably change in the investigated pressure range. Fig. \ref{electronicband} shows a typical picture at 100 GPa. The band structure reveals metallic character with large dispersion bands crossing the Fermi level ($E_{\textrm{Fermi}}$). From Fig. \ref{electronicband}a, we can find that only two bands are partically occupied in the band structure, i.e. two bands across the Fermi level,  referred as Band 1 and Band 2. For $Ccca$-20 phase, G point holds D$_{2h}$ point group. At G point, these two bands across the Fermi level hold B$_{2g}$ and A$_{g}$ symmetry, respectively. The energy bands crossing the Fermi level are depicted in Fig. \ref{electronicband}b. The lower band (Band 1) in energy gives a electron-like Fermi pocket around the G point. Besides, two quasi-parallel pieces of Fermi sheets plot in the Fermi surface present obvious Fermi nesting characteristic, signaling the strong electron-phonon coupling. The Fermi surface originated from Band 2 shows a electron-like characteristic around high symmetry points Z and R in the Brillouin zone. The DOS near the Fermi level is mainly contributed by Sc-3$d$ electrons while Sc-4$s$ electrons make a relatively smaller contribution for the electronic properties of $Ccca$-20.

\begin{figure*}
\epsfig{file=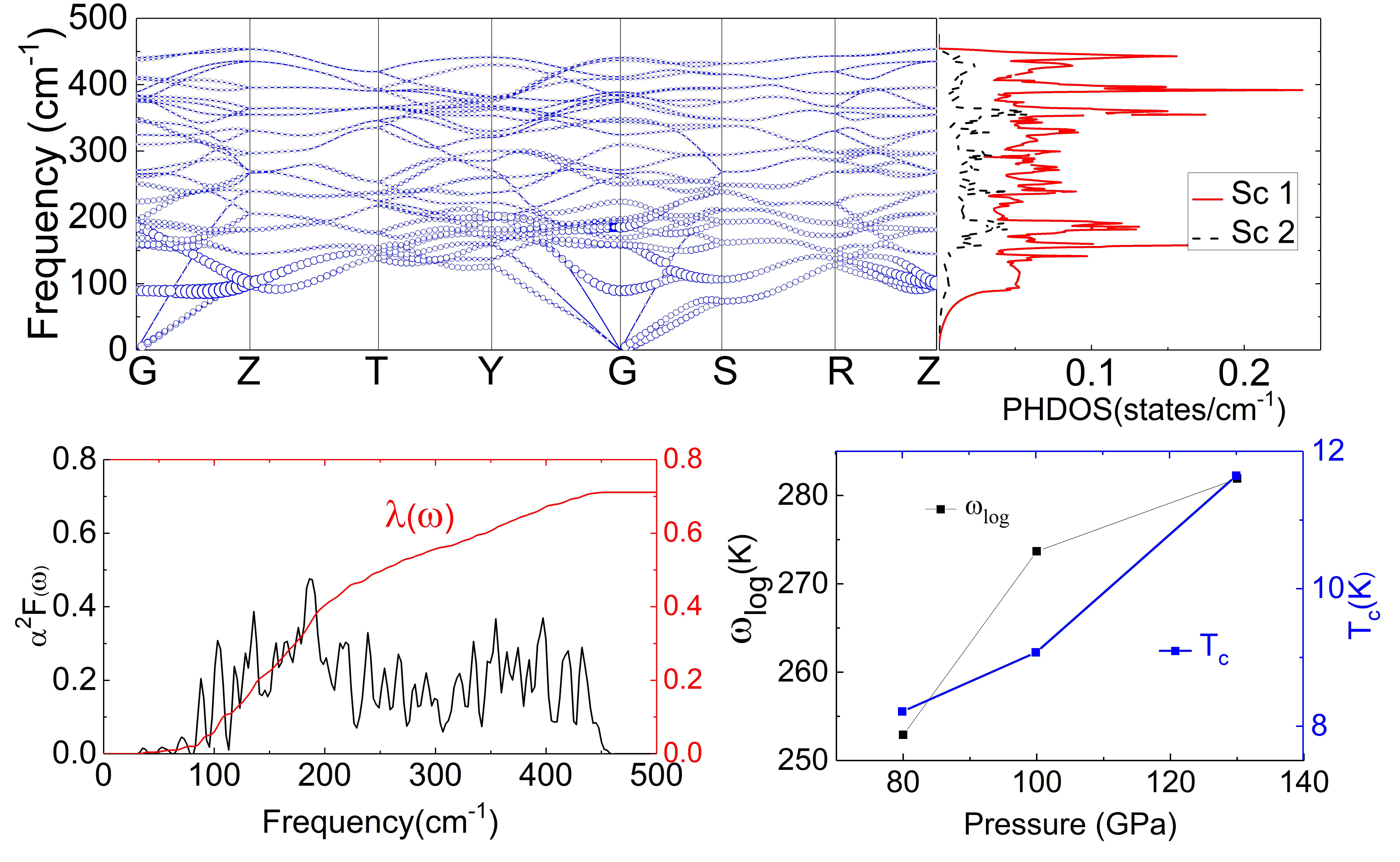, width=0.9\textwidth}
\caption{\label{Phononband} { (a) Phonon dispersion along the high-symmetry directions of the Brillouin zone (left panel) and the partial phonon density of states (PHDOS) (right panel) of the Ccca-Sc at 100 GPa. Blue circles in phonon dispersion show the EPC with the radius proportional to the respective coupling strength. (b) the Eliashberg phonon spectral function and the integrated EPC parameter $\lambda$ as a function of frequency, (c) the superconducting transition temperature $T_\textrm{c}$ and $\omega_\textrm{log}$ as a function of pressure.}}
\end{figure*}

To investigate the possible superconductivity on $Ccca$-20, we also computed the its EPC parameter $\lambda$ and the Eliashberg phonon spectral function $\alpha^2F(\omega)$. The phonon band structure and the projected DOS at 100 GPa are shown in Fig. \ref{Phononband}a. The absence of imaginary frequency modes indicates its dynamic stability. Additional phonon calculations establish the stability range to be between 80 and 130 GPa. A striking feature of the phonon band structure is the presence of soft phonon modes along G-Z, T-Y, and G-S directions, signalling the strong electron-phonon coupling and thus potentially a high $T_\textrm{c}$ value. To quantify the contribution of each phonon branch, we decompose the EPC strength to each q point ($\lambda_q$) along the high symmetry points in the BZ. The sizes of these blue circles in Fig. \ref{Phononband}a indicate their relative contribution to the total $\lambda$. Clearly, we found that the phonons below 240 cm$^{-1}$ contributes significantly to $\lambda$ (see also Fig. \ref{Phononband}b). In particular, the 4th and 7th phonon branches make the largest contributions. By analyzing the their eigenvectors, we found that they are associated with the B$_{1u}$ and B$_{3g}$ vibrational modes. The spectral function $\alpha^2 F(\omega)$ obtained at 100 GPa and the integrated $\lambda$ as a function of frequency are depicted in Fig. \ref{Phononband}b. The results suggest that majority rise of $\lambda$ is in the frequency region between 80-240 cm$^{-1}$, which is consistent with our phonon band analysis. The calculated $\lambda$ is 0.710 at 100 GPa, in which the acoustic modes below 300 cm$^{-1}$ constitutes 78.6 \% of the total $\lambda$, while the higher vibrational modes only contribute 21.4 \%. This result is comparable to the previous studies on other close systems \cite{Chen-PRL-2012}. 

To obtain a rough estimation on the superconducting transition temperature $T_\textrm{c}$, we adopted the modified formula by Allen and Dynes \cite{Allen-PRB-1975}

\begin{equation}
\label{tc}
T_c=\frac{\omega _{\textrm{log}}}{1.2} \exp\bigg[ - \frac{1.04(1+\lambda)}{\lambda-\mu*(1+0.62\lambda)} \bigg],
\end{equation}
where the $\omega_\textrm{log}$ can be calculated directly from the phonon spectrum as follows,
\begin{equation}
\label{omega}
\omega _\textrm{log} = \exp \bigg[ \frac{2}{\lambda} \int_0^\infty \frac{d\omega}{\omega} \alpha^2F(\omega)\ln \omega \bigg].
\end{equation}In the Eq (\ref{tc}), $\mu^*$ is the Coulomb pseudopotential, which is usually between 0.10-0.13 for most metals  \cite{mcmillan1968transition}. At 100 GPa, the calculated $\omega_\textrm{log}$ is 274 K. When $\mu^*$ = 0.11, the estimated $T_\textrm{c}$ is about 9.0 K. To study the pressure dependence of $T_\textrm{c}$, we also performed the EPC calculations at 80 GPa and 130 GPa. The calculated $\omega_\textrm{log}$($\lambda$) at 80, 100 and 130 GPa are 253 K (0.704), 274 K (0.710) and 282 K (0.779), respectively. The $T_\textrm{c}$ of $Ccca$-20 shows a monotonic increase with pressure, from 8.2 K at 80 GPa to 9.0 K at 100 GPa and 11.6 K at 130 GPa. In experiment, the corresponding $T_\textrm{c}$ values were found to be 8 K at 107 GPa and 9 K at 123 GPa \cite{debessai2008comparison-Tc}. We can see that both the tendency and the $T_\textrm{c}$ values are in satisfactory agreement with the experimental data, which supports that the $Ccca$-20 structure is likely to be the experimentally observed Sc-III.

\subsection{PXRD comparison with the previous experimental results}

\begin{figure}[ht]
\epsfig{file=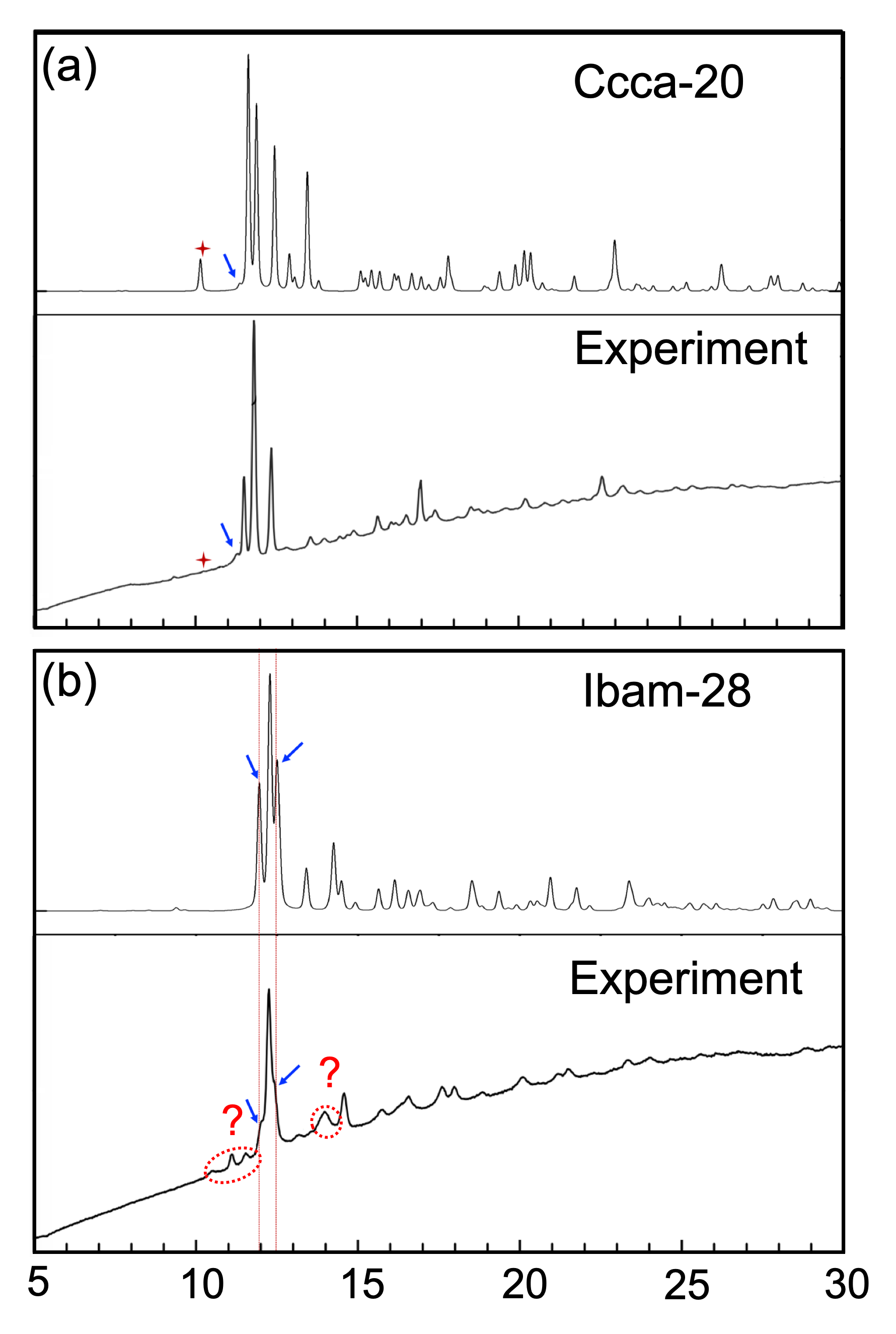, width=0.45\textwidth}
\caption{\label{XRD} {Summary of PXRD comparison between the predicted structures and experimental with a wavelength ($\lambda$) of 0.4428 {\AA}. (a) $Ccca$-20 and Sc-III from experiment \cite{akahama2005new-240GPa}; (b) $Ibam$-28 and Sc-IV from experiment. In general, there is a qualitative agreement between experiment and prediction in terms of the first few strongest peaks. However, the predicted structures exhibit more reflection peaks in high angle range. }}
\end{figure}

In addition to the $T_\textrm{c}$ measurements, another set of available experimental data for Sc-III/IV is the PXRD pattern. Therefore, we also compared the low energy structures with the unsolved experimental Sc-III/IV in terms of the PXRD profiles (see Fig. \ref{XRD}). From Fig. \ref{XRD}a, we found the simulated PXRD of $Ccca$-20 structure share strong similarity with the experimental Sc-III. In particular, both structures have the strongest three peaks at 11.63$^\circ$, 11.88$^\circ$ and 12.42$^\circ$ ($\lambda$ = 0.4428 \AA) at 115 GPa. Another weak peak at 11.36$^\circ$, shown by the blue arrow in Fig. \ref{XRD}b was regarded as the impurity of the sample by Akahama \cite{akahama2005new-240GPa}. But our results suggest that this belongs to an intrinsic reflection peak of $Ccca$-20. Regardless of the qualitative agreement in the peak positions, we fail to find plausible match in peak intensity. This may be due to the possible texturing of the samples used in experiments. Not only changing the intensity, texturing may also diminish some reflection peaks \cite{stassen1996magnetic-texture}. Based on the likely match in PXRD and the energetic stability at the similar pressure range, we suggest that $Ccca$-20 may be the candidate model for Sc-III. 

\begin{figure*}[ht]
\epsfig{file=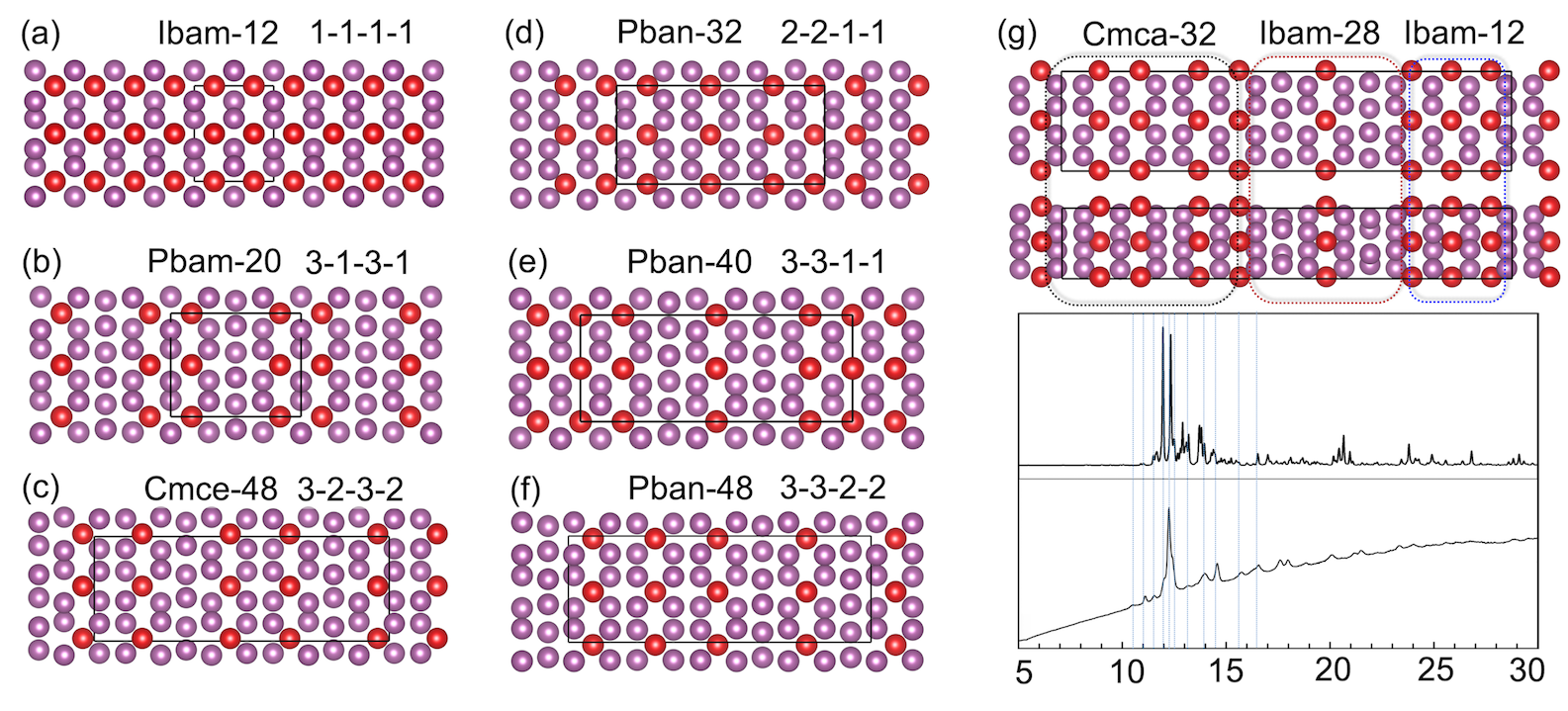, width=0.90\textwidth}
\caption{\label{more} {The atomic structures of (a) $Ibam$-12 (1-1-1-1), (b) $Pbam$-20 (3-1-3-1), (c) $Cmce$-48 (3-2-3-2), (d) $Pban$-32 (2-2-1-1), (e) $Pban$-40 (3-3-1-1) and (f) $Pban$-48 (3-3-2-2). (g) (upper) The atomic structure of a super cell containing different fragments from $Cmca$-32, $Ibam$-28 and $Ibam$-12; (lower) the PXRD profiles between simulation and experiment. The low angle peaks from the supercell structure match well with that from experiment. The Sc atoms in the layer framework are denoted by purple spheres, and Sc atoms in the intercalated layer are denoted by red spheres.}}
\end{figure*}

At higher pressure, our prediction suggested that the $Cmca$-32 is most stable. However, the simulated PXRD of $Cmca$-32 is extremely different from Sc-IV as observed in experiment \cite{akahama2005new-240GPa} (see Fig. S4 \cite{SI}). Another metastable structure $Ibam$-28 seems to provide a better fit to Sc-IV. As shown in Fig. \ref{XRD}b, the experiment PXRD profile has three main peaks, namely two shoulder peaks and one main peak, which can match those in $Ibam$-28 fairly well in terms of both peaks position and intensity. It is reasonable to speculate that $Ibam$-28 may, at least partially explains the observed pattern for Sc-IV. Given that $Ibam$-28 is merely less stable than $Cmca$-32, it is kinetically indeed possible to observe in experiment.

Yet, there are still some weak peaks missing in the simulated PXRD, especially in the low degree (see the dot circles in Fig.\ref{XRD}b). The extra peaks of the experimental PXRD indicate the Sc-IV may be a mixture of different structures. Here, we manually constructed a series of structures with different stacking sequences of A and B layer, and then optimized their geometries at 150 GPa by DFT. Fig. \ref{more} shows these structures (their enthalpies-pressure relations can be found in Fig. S5 \cite{SI}). Based on our earlier descriptions on $Ccca$-20 (2A+1B+2A+1B), $Ibam$-28 (3A+1B+3A+1B) and $Cmca$-32 (2A+1B+1A+1B), we name them as 2-2-2-2, 3-3-3-3 and 2-1-2-1, respectively. Here, the digit number corresponds the number of \textbf{A} layers, and the transverse means the connecting \textbf{B} layer. Following the same convention, we name these trial structures as follows, $Ibam$-12 (1-1-1-1), $Pbam$-20 (3-1-3-1), $Cmce$-48 (3-2-3-2), $Pban$-32 (2-2-1-1), $Pban$-40 (3-3-1-1) and $Pban$-48 (3-3-2-2) and so on. The simulated PXRD profiles of those structures are shown in Fig. S6 \cite{SI}. Comparing with experiment data, we find that the structures with "-1-1" termination indeed exhibit the low angle weak reflection peaks consistent with the experimental pattern. This suggests that the Sc-IV phase may contain small portion of other structures like $Ibam$-12 or other similar structures. To confirm this hypothesis, we constructed a supercell structure which contains $Cmca$-32, $Ibam$-28 and $Ibam$-12 local structure, as shown in Fig.\ref{more}g, the simulated PXRD can indeed match the experiment one well in the entire 2$\theta$ range. Though still speculative, this suggests that the real structure may be described by the structural unit model which has been used to describe the materials grain boundary \cite{Sutton-1989}.

\section{Conclusion}
In conclusion, using \textit{ab initio} evolutionary structure prediction method USPEX, we performed a thorough crystal structure search to explore the high-pressure phases of Sc. We reported $Ccca$-20 structure is likely candidate for the high-pressure allotropes of Sc-III. This is evidence by the qualitative agreement on the PXRD pattern between experiment and theory, and the quantitative agreement on the evolution of superconducting properties. Using $Ccca$-20 as the model system, we found that two partically occupied band across the Fermi level which the low energy band gives a electron-like Fermi pocket around the G point and the high energy present two quasi-parallel pieces of Fermi sheets plot in the Fermi surface, feature as Fermi nesting characteristic. The EPC is mainly contributed by the low frequency phonon modes, signaling the strong electron-phonon coupling. For the high pressure form of Sc-IV, we failed to find any single structure can match the observed PXRD pattern well. Instead, a model based on random stacking of two layered building block seemed to yield the best agreement with the experimental PXRD. This suggested that Sc at high pressure may adopt a complex structure by assembling different structural units. However, the full determination of Sc-III/IV still requires a synergy between experiment and theory. We hope our results here can serve as a guide for following studies in future.

\begin{acknowledgements}
Work at UNLV is supported by the National Nuclear Security Administration under the Stewardship Science Academic Alliances program through DOE Cooperative Agreement DE-NA0001982. S.-C. Zhu is supported by NSFC (Grant No.21703004). Y.-L. Li acknowledges support from the NSFC (Grant No. 11674131) and the 333 project of Jiangsu province. We acknowledge the use of computing resources from XSEDE (TG-DMR180040) and Center for Functional Nanomaterials under contract no. DE-AC02-98CH10086.
\end{acknowledgements}

\bibliography{reference}

\end{document}